\documentclass[english,11pt,a4paper]{article}

\usepackage[english]{babel}
\usepackage[utf8]{inputenc}
\usepackage[a4paper]{geometry}
\usepackage{bm}
\usepackage{amsmath}
\usepackage{amsthm}
\usepackage{amssymb}
\usepackage{cite}

\usepackage{mathdots}
\usepackage{setspace}
\date{}

\usepackage{authblk}

\usepackage[bookmarks=false,breaklinks=false,pdfborder={0 0 1},backref=false,colorlinks=true]{hyperref}
\hypersetup{allcolors=blue}

\author[$\dagger$]{Maciej Błaszak}
\author[$\ddagger$]{Krzysztof Marciniak}
\author[$\dagger$]{Błażej M. Szablikowski\footnote{Corresponding author.}}

\affil[$\dagger$]{ISQI, Faculty of Physics and Astronomy, Adam Mickiewicz University\authorcr
Uniwersytetu Pozna\'{n}skiego 2, 61-614 Poznań, Poland\authorcr
\texttt{blaszakm@amu.edu.pl, bszablik@amu.edu.pl}\authorcr\mbox{}}

\affil[$\ddagger$]{Department of Science and Technology\authorcr
Link\"{o}ping University, Campus Norrk\"{o}ping\authorcr
601 74 Norrk\"{o}ping, Sweden\authorcr
\texttt{krzma@itn.liu.se}}

\newtheorem{theorem}{Theorem}

\newtheorem{example}[theorem]{Example}

\begin{document}

\title{\textsc{Stationary systems of the AKNS hierarchy}}


\maketitle

\begin{abstract}
In this article we investigate stationary systems of the AKNS hierarchy. We
prove that all these systems are classical separable Stäckel systems. The
explicit transformation from jet variables to separation coordinates is obtained by comparing
Lax representations in jet and separable coordinates.

\end{abstract}


\section{Introduction}

From the classical works of Bogoyavlensky, Novikov \cite{novik76} and Mokhov
\cite{mokhov} it is well known that stationary flows of soliton hierarchies
are finite dimensional Hamiltonian systems. Up till now the research was
focused on stationary flows of the KdV hierarchy and its generalization, i.e.
coupled KdV hierarchies. It was proved that they are bi-Hamiltonian
\cite{ant87,ant92,FH} and separable according to the Hamilton-Jacobi theory
\cite{1,2}. In fact, in \cite{1,2} we investigated not only stationary flows but 
the so-called stationary systems, that is finite dimensional systems of evolutionary 
equations of the form
\begin{equation}\label{statsysI}%
u_{t_{1}}=\bm{K}_{1}[u],\qquad 
u_{t_{2}}=\bm{K}_{2}[u],\qquad
\ldots,\qquad 
u_{t_{n}}=\bm{K}_{n},[u]\qquad \bm{K}_{n+1}[u]=0, 
\end{equation}
related to the infinite hierarchy $u_{t_{i}}=\bm{K}_{i}[u]$, $i\in\mathbb{N}$, of commuting 
evolutionary-type systems. In this setting, the last equation in \eqref{statsysI} 
is the stationary flow of the infinite hierarchy that defines a finite-dimensional 
manifold (stationary manifold) in the infinite-dimensional jet space of variables $u,u_x,u_{xx},\ldots$. 
This manifold is invariant with respect to all the remaining evolution equations in \eqref{statsysI}. 
In \cite{1,2} we proved that stationary KdV and cKdV systems are in fact Stäckel separable systems
and proved that they are multi-Hamiltonian.

In this article we thoroughly analyse stationary systems of the Ablowitz-Kaup-Newell-Segur (AKNS) hierarchy \cite{AKNSc}
and prove their separability. These systems have not been investigated prior to this article. What is interesting,
the constructed finite dimensional systems are not multi-Hamiltonian, contrary to
the KdV and cKdV cases described in \cite{1} and in \cite{2}, respectively. 
This is due to the fact that the AKNS hierarchy itself has only one local invariant Poisson tensor.

In Sections \ref{s1} and \ref{s15} we remind the construction of the AKNS hierarchy and its
Lax and zero-curvature representations, respectively. In Section \ref{s2} we present the
stationary restrictions of the AKNS hierarchy and the related restrictions of its Lax
and zero-curvature representations. In consequence, we receive spectral
(separation) curves that will be related with Stäckel separable systems. 
In order to keep the article self-contained, in Section \eqref{s3} we briefly remind 
known facts about Stäckel systems and their Lax
representations. In Section \ref{s4}, we relate the Stäckel
systems constructed in Section \ref{s3} with AKNS stationary systems,
constructed in Section \ref{s2}, through an explicit transformation between
separation coordinates and jet coordinates. This way we prove that stationary
systems of the AKNS hierarchy are Stäckel systems.

\section{The AKNS hierarchy \label{s1}}

Let us remind some important facts about the AKNS hierarchy \cite{AKNSc,Ab,N} in real representation. 
It is a bi-Hamiltonian hierarchy 
\begin{equation}
\begin{pmatrix}
  q\\ r
\end{pmatrix}_{t_n}
=\bm{K}_{n}=\pi_{0}\gamma_{n}=\pi_{1}\gamma_{n-1},\qquad n=1,2,\ldots, \label{AKNS}%
\end{equation}
where $q=q(x,t_1,t_2,\ldots)$ and $r=r(x,t_1,t_{2},\ldots)$ are field variables depending
on the spatial variable $x$ and on evolution parameters (times) $t_n$. This hierarchy is defined
on a (functional) infinite-dimensional manifold $\mathcal{M}$ with jet coordinates
$q,r,q_x,r_x,q_{xx},r_{xx}\ldots$. In \eqref{AKNS} $\gamma_{n}$ are exact one-forms while
\[
\pi_{0}=\begin{pmatrix}
  0 & 1\\
-1 & 0
\end{pmatrix},\qquad 
\pi_{1}=
\begin{pmatrix}
-q\partial_{x}^{-1}q & -\frac{1}{2}\partial_{x}+q\partial_{x}^{-1}r\\
-\frac{1}{2}\partial_{x}+r\partial_{x}^{-1}q & -r\partial_{x}^{-1}r
\end{pmatrix}
\]
are two compatible Poisson tensors. Here and in what follows
\[
 \partial_x^{-1} \equiv \int_{-\infty}^x dx.
\]
The hierarchy is constructed by means of the recursion operator
\begin{equation}\label{recop}
\mathcal{R}:=\pi_{1}\pi_{0}^{-1}=
\begin{pmatrix}
-\frac{1}{2}\partial_{x}+q\partial_{x}^{-1}r & q\partial_{x}^{-1}q\\
-r\partial_{x}^{-1}r & \frac{1}{2}\partial_{x}-r\partial_{x}^{-1}q
\end{pmatrix}
\end{equation}
as follows%
\begin{equation}
\bm{K}_{n}=\mathcal{R}^{n-1}\bm{K}_{1},\qquad n=1,2,\ldots\,.\label{rec}%
\end{equation}
By the hereditary property of the recursion operator \eqref{recop} all vector fields $\bm{K}_n$ mutually commute and all
one-forms (co-symmetries) $\gamma_n$ are closed, thus there exists an infinite sequence of
functionals $H_{n}=\int h_{n}dx$ such that $\gamma_n=\delta H_{n}$.
In the recursion \eqref{rec} all the integration constants vanish if, for instance,
we assume the following boundary conditions  
\begin{equation}\label{bound}
  \lim_{x\rightarrow -\infty} q,r,q_x,r_x,\ldots = 0.
\end{equation}

In particular, the first few symmetries have the form
\[
\bm{K}_1 = \begin{pmatrix}
  -2q\\ 2r
\end{pmatrix},\quad
\bm{K}_2 = \begin{pmatrix}
  q_{x}\\ r_{x}
\end{pmatrix},\quad
\bm{K}_3 = \begin{pmatrix}
  -\frac{1}{2}q_{xx}+q^{2}r\\ \frac{1}{2}r_{xx}-qr^{2}
\end{pmatrix},\quad
\bm{K}_4 = \begin{pmatrix}
  \frac{1}{4}q_{xxx}-\frac{3}{2}qrq_{x}\\ \frac{1}{4}r_{xxx}-\frac{3}{2}qrr_{x}
\end{pmatrix},
\]
first few co-symmetries are
\[
\gamma_1 = \begin{pmatrix}
  -2r\\ -2q
\end{pmatrix},\quad
\gamma_2 = \begin{pmatrix}
  -r_x\\ q_{x}
\end{pmatrix},\quad
\gamma_3 = \begin{pmatrix}
  -\frac{1}{2}r_{xx}+qr^{2}\\ -\frac{1}{2}q_{xx}+q^{2}r
\end{pmatrix},\quad
\gamma_4 = \begin{pmatrix}
  -\frac{1}{4}r_{xxx}+\frac{3}{2}qrr_{x}\\ \frac{1}{4}q_{xxx}-\frac{3}{2}qrq_{x}
\end{pmatrix},
\]
while first Hamiltonian densities are
\[
h_{1}=-2qr,\qquad h_{2}=rq_{x},\qquad 
h_{3}=\frac{1}{2}q_{x}r_{x}+\frac{1}{2}q^{2}r^{2},\qquad 
h_{4}=\frac{1}{4}rq_{xxx}+\frac{3}{4}q^{2}rr_{x}.
\]%

\section{Lax representation of the AKNS hierarchy \label{s15}}

The hierarchy of vector fields \eqref{AKNS} can also be reconstructed form the
so-called zero-curvature equations. Consider the following linear
problem%
\begin{align}\label{laxpair}
\Psi_{x}=\mathbb{L}\Psi,\qquad \Psi_{t_{k}}=\mathbb{U}_{k}\Psi,\qquad k=1,2,\ldots,
\end{align}
where $\mathbb{L}$, $\mathbb{U}$ are some $2\times2$ matrices depending on the isospectral 
parameter~$\lambda$, i.e. 
$\lambda_{t_{n}}=0$ for all $n$, and jet coordinates on $\mathcal{M}$. The compatibility
conditions (that is the conditions for existence of a common multi-time
solution $\Psi(x,t_{1},t_{2},\ldots)$) for this problem are%
\begin{align}\label{cnd}
(\Psi_{x})_{t_{n}} =(\Psi_{t_{n}})_{x},\qquad 
(\Psi_{t_m})_{t_n} =(\Psi_{t_n})_{t_m},\qquad m,n=1,2,\ldots\,.
\end{align}
The first set of conditions in \eqref{cnd} is equivalent to the Lax equations 
\begin{equation}
\mathbb{L}_{t_{n}}=\left[  \mathbb{U}_n,\mathbb{L}\right]  +\left(  \mathbb{U}_{n}\right)_{x},\qquad n=1,2,\ldots,\label{Lax}%
\end{equation}
while the remaining conditions in \eqref{cnd} lead to the zero-curvature equations
\begin{equation}
\left(  \mathbb{U}_{m}\right)  _{t_{n}}-\left(  \mathbb{U}_{n}\right)  _{t_{m}}%
+[\mathbb{U}_{m},\mathbb{U}_{n}]=0,\qquad m,n=1,2,\ldots,\label{zero}%
\end{equation}
which are equivalent to the commutativity of vector fields associated with \eqref{Lax}.

For the AKNS hierarchy the Lax matrix $\mathbb{L}$ is of the form%
\[
\mathbb{L}=%
\begin{pmatrix}
-\lambda & q\\
r & \lambda
\end{pmatrix}
\]
and the auxiliary matrices $U_{n}$ are given by%
\begin{equation}\label{aux}
  \mathbb{U}_n =
  \begin{pmatrix}
    \mathbb{A}_n & \mathbb{B}_n\\ \mathbb{C}_n & -\mathbb{A}_n
  \end{pmatrix},\qquad n=1,2,\ldots\,,
\end{equation}
with the coefficients 
\begin{align*}
  \mathbb{A}_n = -\lambda^{n-1} + \frac{1}{2}\sum_{i=2}^{n-1}a_i\lambda^{n-i-1},\qquad 
   \mathbb{B}_n = -\frac{1}{2}\sum_{i=1}^{n-1}b_i\lambda^{n-i-1},\qquad \mathbb{C}_n = \frac{1}{2}\sum_{i=1}^{n-1}c_i\lambda^{n-i-1}.
\end{align*}
Then, by the Lax equations \eqref{Lax} we obtain, for $n\geqslant 1$, the following
equalities  
\begin{align}\label{eqs}
  q_{t_n} = 2 q \mathbb{A}_n + 2 \mathbb{B}_n \lambda + (\mathbb{B}_n)_x,\qquad
  r_{t_n} = -2r \mathbb{A}_n - 2 \mathbb{C}_n \lambda + (\mathbb{C}_n)_x, 
\end{align}
together with the conditions
\begin{equation}\label{cond}
  (\mathbb{A}_n)_x = q \mathbb{C}_n - r \mathbb{B}_n .
\end{equation}
Requiring consistency of the equalities \eqref{eqs} and \eqref{cond} we find that for arbitrary
$n\geqslant 1$
\begin{align}\label{eqa}
  b_n = -\frac{1}{2}(b_{n-1})_x + q a_{n-1},\qquad c_n = \frac{1}{2}(c_{n-1})_x - r a_{n-1}
\end{align}
and
\begin{equation}\label{eqb}
    (a_n)_x = r b_n + qc_n .
\end{equation}
For simplicity, above and further we assume that $a_0 = -2$ and $a_1=b_0=c_0=0$.
Hence, for $n\geqslant 1$
\[
  q_{t_n} = q a_{n-1}  - \frac{1}{2}(b_{n-1})_x \equiv b_n,\qquad r_{t_n} = -r a_{n-1}  + \frac{1}{2}(c_{n-1})_x \equiv c_n
\]
and consequently 
\begin{equation}\label{K12}
  b_n = (K_n)_1,\qquad c_n = (K_n)_2,\qquad a_n = \partial_x^{-1}\left ( r (K_n)_1 + q (K_n)_2\right).
\end{equation}
This provides the standard Lax representation of the AKNS hierarchy \eqref{AKNS}.

In particular we have
\[
\mathbb{U}_{1} = \begin{pmatrix}
-1 & 0\\
0 & 1
\end{pmatrix},\qquad
 \mathbb{U}_{2}=\begin{pmatrix}
-\lambda & q\\
r & \lambda
\end{pmatrix}\equiv \mathbb{L},\qquad%
 \mathbb{U}_{3}= \begin{pmatrix}
-\lambda^{2}+\frac{1}{2}qr & q\lambda-\frac{1}{2}q_{x}\\
r\lambda+\frac{1}{2}r_{x} & \lambda^{2}-\frac{1}{2}qr
\end{pmatrix},
\]
\[
\mathbb{U}_{4} =\begin{pmatrix}
-\lambda^{3}+\frac{1}{2}qr\lambda-\frac{1}{4}rq_{x}+\frac{1}{4}qr_{x} &
q\lambda^{2}-\frac{1}{2}q_{x}\lambda+\frac{1}{4}q_{xx}-\frac{1}{2}q^{2}r\\
r\lambda^{2}+\frac{1}{2}r_{x}\lambda+\frac{1}{4}r_{xx}-\frac{1}{2}qr^{2} &
\lambda^{3}-\frac{1}{2}qr\lambda+\frac{1}{4}rq_{x}-\frac{1}{4}qr_{x}
\end{pmatrix}.
\]

Alternatively, the auxiliary matrices $\mathbb{U}_n$ can be defined by the following projections
\begin{equation}\label{Un}
  \mathbb{U}_n = \left [\lambda^{n-1}\mathcal{U}\right]_+,\qquad n=1,2,\ldots,
\end{equation}
where $[\cdot]_+$ means the projection on the polynomial part in $\lambda$
and
\begin{equation}\label{gen}
  \mathcal{U} =
  \begin{pmatrix}
    \mathcal{A} & \mathcal{B}\\ \mathcal{C} & -\mathcal{A}
  \end{pmatrix}
\end{equation}
is a generating matrix, with coefficients being series of the form
\begin{align}\label{series}
  \mathcal{A} = -1 + \frac{1}{2}\sum_{i=2}^{\infty}a_i\lambda^{-i},\qquad 
   \mathcal{B} = -\frac{1}{2}\sum_{i=1}^{\infty}b_i\lambda^{-i},\qquad \mathcal{C} = \frac{1}{2}\sum_{i=1}^{\infty}c_i\lambda^{-i}.
\end{align}
Due to \eqref{Lax} the matrix $\mathcal{U}$ must satisfy the condition
\begin{equation}\label{cond2}
  \left [\mathcal{U},L\right] + \mathcal{U}_x = 0.
\end{equation}
By \eqref{cond2} we find that 
\begin{align}\label{eqsa}
  2 q \mathcal{A} + 2 \mathcal{B} \lambda + (\mathcal{B})_x = 0,\qquad
  -2r \mathcal{A} - 2 \mathcal{C} \lambda + (\mathcal{C})_x = 0
\end{align}
together with 
\begin{equation}\label{eqsb}
  (\mathcal{A})_x = q \mathcal{C} - r \mathcal{B} .
\end{equation}
Substituting \eqref{series} to \eqref{eqsa} and \eqref{eqsb} we obtain again \eqref{eqa} and \eqref{eqb}.
The construction of the Lax representation for the AKNS hierarchy \eqref{AKNS}
by means of \eqref{Un} and \eqref{cond2}, in fact, coincides with the classical $r$-matrix
approach to AKNS hierarchy, see for instance \cite{BSz} and references therein.

From the point of view of further considerations the important result is that the determinant of the generating matrix~\eqref{gen} is constant, that is 
\[
  \det\mathcal{U} \equiv -\mathcal{A}^2 - \mathcal{B}\mathcal{C} = -1 + \sum_{i=1}^{\infty}\varepsilon_i\lambda^{-i},\qquad \varepsilon_i \in\mathbb{R}.
\]
This follows by differentiating $\det\mathcal{U}$ with respect to the spatial variable $x$
and taking into account \eqref{eqsa} and \eqref{eqsb}, that is 
\[
  (\det U)_x = -\left( \mathcal{A}^2 + \mathcal{B}\mathcal{C}\right )_x  
  = 0.
\]
Now, by the boundary conditions \eqref{bound} we see that $\varepsilon_i=0$.
Hence, we have the additional general condition
\[
  \mathcal{A}^2 + \mathcal{B}\mathcal{C} = 1,
\]
from which we obtain the following algebraic recursion on the coefficients $a_n$,
\begin{equation}\label{eqc}
  a_n = \frac{1}{4}\sum_{i=2}^{n-2}a_i a_{n-i} - \frac{1}{4} \sum_{i=1}^{n-1}b_i c_{n-i},\qquad n=2,3,\ldots.
\end{equation}
This means that the AKNS hierarchy \eqref{AKNS} can be obtained purely by algebraic recursion given
by \eqref{eqa} and \eqref{eqc}. This means that the integration in the recursion \eqref{rec}, or equivalently in \eqref{eqb}, can be avoided. This also shows that the AKNS hierarchy consists of local vector fields.

Next, for further considerations, we need to slightly reformulate the AKNS Lax representation.
By \eqref{zero}, since $(\mathbb{U}_1)_{t_k}=0$ for all $k\geqslant 1$,  
we have that
\[
  \left(\mathbb{U}_{k}\right)_{t_{1}} = [\mathbb{U}_{1},\mathbb{U}_{k}]
\]
and in consequence
\begin{equation}\label{abc}
  \dot{\mathbb{A}}_k = 0,\qquad \dot{\mathbb{B}}_k = -2\mathbb{B}_k,\qquad
   \dot{\mathbb{C}}_k = 2 \mathbb{C}_k.
\end{equation}
Here and in what follows, the dot means the derivative with respect
to the first evolution parameter $t_1$, i.e.~$\dot{\xi}\equiv\xi_{t_{1}}$.
Let
\[
  \mathbb{P}_k :=  -2\mathbb{A}_k + \mathbb{B}_k-\mathbb{C}_k 
 =   2\lambda^{k-1} + \frac{1}{2}\sum_{i=1}^{k-1}P_i\lambda^{k-i-1},
\]
where
\[
  P_i = -a_i - \frac{1}{2}b_i - \frac{1}{2}c_i,\qquad i=1,2,\ldots. 
\]
Then, by \eqref{abc} we see that
\[
  \dot{\mathbb{P}}_k = -2\mathbb{B}_k-2\mathbb{C}_k,\qquad 
  \ddot{\mathbb{P}}_k = 4\mathbb{B}_k-4\mathbb{C}_k = 4\mathbb{P}_k+8\mathbb{A}_k,\qquad
  \dddot{\mathbb{P}}_k = 4\dot{\mathbb{P}}_k
\]
and thus
\[
  \mathbb{A}_k = \frac{1}{8}\ddot{\mathbb{P}}_k -\frac{1}{2}\mathbb{P}_k,\qquad
  \mathbb{B}_k = \frac{1}{8}\ddot{\mathbb{P}}_k -\frac{1}{4}\dot{\mathbb{P}}_k,\qquad
  \mathbb{C}_k = -\frac{1}{8}\ddot{\mathbb{P}}_k -\frac{1}{4}\dot{\mathbb{P}}_k.
\]
Hence, we can rewrite the auxiliary matrices \eqref{aux} in the form 
\begin{equation}\label{UkP}
  \mathbb{U}_k   =\frac{1}{8}\begin{pmatrix}
   \ddot{\mathbb{P}}_k -4\mathbb{P}_k  & \ddot{\mathbb{P}}_k -2\dot{\mathbb{P}}_k\\ 
   -\ddot{\mathbb{P}}_k -2\dot{\mathbb{P}}_k & -\ddot{\mathbb{P}}_k +4\mathbb{P}_k
  \end{pmatrix},\qquad k=1,2,\ldots\,.
\end{equation}


\section{Stationary systems \label{s2}}

The $(n+1)$-th stationary flow of the AKNS hierarchy is defined by:
\begin{equation}
\begin{pmatrix}q\\ r \end{pmatrix}_{t_{n+1}}
=0\qquad\text{or equivalently}\qquad \bm{K}_{n+1}=0. \label{statflow}%
\end{equation}
The stationary flow \eqref{statflow} can be obtained by imposing the following
constraint
\begin{equation}\label{constr}
\Psi_{t_{n+1}}=\mu\Psi,\qquad \text{or equivalently}\qquad \mathbb{U}_{n+1}\Psi=\mu\Psi,
\end{equation}
on the linear problems \eqref{laxpair},
where $\mu$ is another isospectral parameter. Indeed, the first constraint in \eqref{constr}
and the compatibility condition $(\Psi_{x})_{t_{n+1}}=(\Psi_{t_{n+1}})_{x}$
gives
\begin{equation}
\frac{d}{d t_{n+1}}\mathbb{L}=0, \label{constrcc}%
\end{equation}
which is equivalent to \eqref{statflow}. Alternatively, the compatibility
condition between the second condition in \eqref{constr} and $\Psi_{x}=\mathbb{L}\Psi$ yields
the equation
\[
\frac{d}{dx}\mathbb{U}_{n+1}=[\mathbb{L},\mathbb{U}_{n+1}],%
\]
equivalent with \eqref{constrcc} by the Lax equation \eqref{Lax}
valid under the constraint \eqref{constr}.

The vector field $\bm{K}_{n+1}$ and the matrix $\mathbb{U}_{n+1}$ depend on $2n+2$ jet variables:
$q,r,q_x,r_x,\ldots,\linebreak[0] q_{nx},r_{nx}$. Thus, the stationary restriction
\eqref{statflow} provides constraint on the infinite-dimensio\-nal (functional)
manifold $\mathcal{M}$, on which the AKNS hierarchy is defined, reducing it to the
finite-dimensional (stationary) submanifold $\mathcal{M}_{n}$ of dimension
$2n$. Using \eqref{statflow} and its differential consequences we can eliminate all
terms of order $n$ and higher. Thus, the coordinates on the stationary
manifold $\mathcal{M}_{n}$ are provided by the jet coordinates: $q,r,q_x,r_x,\ldots,\linebreak[0] q_{(n-1)x},r_{(n-1)x}$. 
Since all the vector fields  $\bm{K}_{i}$ of the AKNS hierarchy \eqref{AKNS} pairwise commute, the stationary manifold $\mathcal{M}_{n}$ is invariant 
with respect to all $\bm{K}_{i}$. Thus, it makes sense to consider the following finite-dimensional system:
\begin{equation}\label{statsys}%
\begin{pmatrix}q\\ r \end{pmatrix}_{t_{1}}=\bm{K}_{1},\qquad 
\begin{pmatrix}q\\ r \end{pmatrix}_{t_{2}}=\bm{K}_{2},\qquad
\ldots,\qquad 
\begin{pmatrix}q\\ r \end{pmatrix}_{t_{n}}=\bm{K}_{n},\qquad \bm{K}_{n+1}=0, 
\end{equation}
which we will further refer to as the \emph{$n$-th stationary AKNS system}. 
This is the object of study of the present article.

The finite hierarchy of Lax equations associated to \eqref{statsys} is given by
\begin{equation}
(\mathbb{U}_{n+1})_{t_k}=[\mathbb{U}_{k},\mathbb{U}_{n+1}],\qquad k=1,2,\ldots,n,\label{Laxs}%
\end{equation}
where elements of $\mathbb{U}_{k}$ are now constrained by the last equation in \eqref{statsys}. In fact, the Lax equations \eqref{Laxs} are a consequence of the compatibility between linear problems \eqref{constr} and \eqref{laxpair}. Alternatively, \eqref{Laxs} follow from the zero-curvature equations \eqref{zero}
and the fact that $(\mathbb{U}_{k})_{t_{n+1}}=0$, which now is a consequence of the compatibility between the former constraint \eqref{constr} and the later linear problem \eqref{laxpair}.  
Notice that for the stationary system \eqref{statsys} the matrix $\mathbb{U}_{n+1}$ plays the role of the Lax matrix.

By the Lax equations \eqref{Laxs} we obtain for $k=1,\ldots,n$ the following equalities
\begin{equation}\label{seq}
\left(\mathbb{B}_{n+1}\right)_{t_k} = 
2 \mathbb{A}_k  \mathbb{B}_{n+1} -2 \mathbb{A}_{n+1} \mathbb{B}_k,\qquad
\left(\mathbb{C}_{n+1}\right)_{t_k} = 
-2 \mathbb{A}_k  \mathbb{C}_{n+1} +2 \mathbb{A}_{n+1} \mathbb{C}_k
\end{equation}
and
\begin{equation}\label{seqc}
\left(\mathbb{A}_{n+1}\right)_{t_k} = \mathbb{B}_k \mathbb{C}_{n+1}-\mathbb{B}_{n+1} \mathbb{C}_k. 
\end{equation}
From \eqref{seq} and \eqref{seqc} it immediately follows that on the stationary manifold 
$\mathcal{M}_n$ defined by \eqref{statflow}, or equivalently \eqref{constr}, 
the time derivatives of the determinant of $U_{n+1}$ vanish, that is
\begin{equation}\label{pU}
  (\det \mathbb{U}_{n+1})_{t_k} = -\left( \mathbb{A}_{n+1}^2 + \mathbb{B}_{n+1}\mathbb{C}_{n+1}\right )_{t_k} = 0,
\end{equation}
for $k=1,\ldots, n$. Moreover, defining $h_k$ as the coefficients of the polynomial $\det \mathbb{U}_{n+1}$
\begin{equation}\label{dUn}
  \det \mathbb{U}_{n+1} = -\mathbb{A}_{n+1}^2 - \mathbb{B}_{n+1}\mathbb{C}_{n+1} =: 
  \sum_{k=0}^{2n}h_{k}\lambda^{k},
\end{equation}
we see that $h_{2n}=-1$, while the coefficients $h_k$ for $k=n,\ldots,2n-1$ identically vanish by \eqref{eqc}:
\[
h_k = a_{2n-k}-\frac{1}{4}\sum_{i=2}^{2n-k-2}a_i a_{2n-i-k} + \frac{1}{4} \sum_{i=1}^{2n-k-1}b_i c_{2n-i-k}
  = 0,
\]
while for the remaining $k$ we obtain
\begin{align}\label{hk}
  h_k &= \frac{1}{4}\sum_{i=n-k}^{n}\left (b_i c_{2n-i-k}- a_i a_{2n-i-k}\right),\quad
  \text{here}\quad k=0,\ldots,n-1. 
\end{align}
By \eqref{pU} this means that \eqref{hk} form an $n$-element set of integrals of motion for the
$n$-th stationary AKNS system \eqref{statsys}. In fact, we are going to show that they are functionally independent Hamiltonians of the respective flows in \eqref{statsys}.   

Further, the constraint in \eqref{constr} has non-trivial solutions provided that
\begin{equation}
\det\left(  \mathbb{U}_{n+1}-\mu\mathbb{I}\right)  =0,\label{L11}%
\end{equation}
which takes the following explicit form 
\[
  \mathbb{A}_{n+1}^2 + \mathbb{B}_{n+1}\mathbb{C}_{n+1} = \mu^2.
\]
Hence, comparing it with \eqref{dUn}, the equation \eqref{L11} determines the following spectral curve
\begin{equation}
\lambda^{2n}-\sum_{k=0}^{n-1}h_{k}\lambda^{k}=\mu^{2}, \label{L13}%
\end{equation}
which determines a common level for the constants of
motion \eqref{hk} of the respective $n$-th stationary AKNS system \eqref{statsys}. 

If we consider the alternative representation of $\mathbb{U}_{n+1}$ given by \eqref{UkP}, then
the characteristic equation \eqref{L11} takes the form
\[
-\frac{1}{8}\mathbb{P}_{n+1}\ddot{\mathbb{P}}_{n+1}+\frac{1}{16}\dot{\mathbb{P}}_{n+1}^{\,2}
+\frac{1}{4}\mathbb{P}_{n+1}^{\,2}=\mu^{2},
\]
(cf. (61) in \cite{2} and (25) in \cite{1}).\footnote{Notice that the related spectral curves in
\cite{1,2} are defined with respect to the spatial variable $x$ that can be identified with
the evolution parameter $t_1$. In the case of the AKNS hierarchy \eqref{AKNS} we cannot make such identification
as the translational symmetry is only the second symmetry in the hierarchy.}

\begin{example}\label{ex1} 
For $n=2$ the stationary AKNS system \eqref{statsys}
consists of two evolution equations
\[
 \begin{pmatrix}q\\ r \end{pmatrix}_{t_{1}} = 
 \begin{pmatrix}
  -2q\\ 2r
\end{pmatrix}\equiv  \bm{K}_1,\qquad
\begin{pmatrix}q\\ r \end{pmatrix}_{t_{2}} = 
\begin{pmatrix}
  q_{x}\\ r_{x}
\end{pmatrix}\equiv  \bm{K}_2
\]
and of the stationary flow 
\[
 \bm{K}_3\equiv \begin{pmatrix}
  -\frac{1}{2}q_{xx}+q^{2}r\\ \frac{1}{2}r_{xx}-qr^{2}
\end{pmatrix} =
\begin{pmatrix}0\\ 0 \end{pmatrix}.
\]
The associated spectral curve \eqref{L13} is
\[
\lambda^4- h_1 \lambda - h_0 =\mu^{2},%
\]
where the two constants of motion \eqref{hk} are
\begin{equation}
h_{0}=\frac{1}{4}q_{x}r_{x}-\frac{1}{4}q^{2}r^{2},\qquad 
h_{1}=\frac{1}{2}rq_{x}-\frac{1}{2}qr_{x}.  \label{hn2}%
\end{equation}
\end{example}

\begin{example}\label{ex2}
For $n=3$ the stationary system \eqref{statsys}
consists of three evolution equations
\[
 \begin{pmatrix}q\\ r \end{pmatrix}_{t_{1}} = 
 \begin{pmatrix}
  -2q\\ 2r
\end{pmatrix}\equiv  \bm{K}_1,\quad
\begin{pmatrix}q\\ r \end{pmatrix}_{t_{2}} = 
\begin{pmatrix}
  q_{x}\\ r_{x}
\end{pmatrix}\equiv  \bm{K}_2,\quad
\begin{pmatrix}q\\ r \end{pmatrix}_{t_{3}} =  \begin{pmatrix}
  -\frac{1}{2}q_{xx}+q^{2}r\\ \frac{1}{2}r_{xx}-qr^{2}
\end{pmatrix}\equiv\bm{K}_3
\]
and of the stationary flow 
\[
  \bm{K}_4 \equiv \begin{pmatrix}
  \frac{1}{4}q_{xxx}-\frac{3}{2}qrq_{x}\\ \frac{1}{4}r_{xxx}-\frac{3}{2}qrr_{x}
\end{pmatrix}=
\begin{pmatrix}0\\ 0 \end{pmatrix}.
\]
The related spectral curve \eqref{L13} is
\[
\lambda^6 -h_2\lambda^2 - h_1 \lambda - h_0 =\mu^{2},%
\]
where
\begin{equation}\label{hn3}
  \begin{split}
h_{0}  &  =-\frac{1}{16}q_{xx}r_{xx}+\frac{1}{8}qr^{2}q_{xx}+
\frac{1}{8}q^{2}rr_{xx}+\frac{1}{8} q r q_x r_x-\frac{1}{16}q^{2}r_{x}^{2}-\frac{1}{16}r^{2}q_{x}^{2}
-\frac{1}{4}q^3r^3,\\
h_{1}  &  =\frac{1}{8}q_{x}r_{xx}-\frac{1}{8}r_{x}q_{xx},\qquad 
h_{2}  =-\frac{1}{4}qr_{xx}-\frac{1}{4}rq_{xx}+\frac{1}{4}q_{x}r_{x}
+\frac{3}{4}q^{2}r^{2}.
    \end{split}
\end{equation}
\end{example}

\section{Stäckel systems generated by spectral curves \label{s3}}

Consider the algebraic curve
\begin{equation}
\lambda^{2n}+\sum_{k=1}^{n}H_{k}\lambda^{n-k}=\mu^{2},\qquad n\in\mathbb{N}  \label{sc}%
\end{equation}
on the $(\lambda,\mu)$-plane (after the, so far, formal identification $H_k\equiv -h_{n-k}$ this curve coincides with the spectral curve \eqref{L13}). This is a separation curve that is a source of finite-dimensional Stäckel systems of Benenti type. Let us briefly remind this construction along with basic facts about Stäckel systems.
For a fixed $n\in\mathbb{N}$, consider a $2n$-dimensional
Poisson manifold $(M=T^{\ast}Q,\pi)$ and a particular set $\bm{\xi} =(\bm{\lambda},\bm{\mu})$ of Darboux
(canonical) coordinates, where $\bm{\lambda}=(\lambda_{1},\ldots,\lambda_{n})^{T}$
are coordinates on an $n$-dimensional configuration space $Q$ and
$\bm{\mu}=(\mu_{1},\ldots,\mu_{n})^{T}$ are the fibre (momenta)
coordinates, i.e.~$\{\lambda_{i},\lambda_{j}\}_{\pi}
=\{\mu_{i},\mu_{j}\}_{\pi}=0$, $\{\lambda_{i},\mu_{j}\}_{\pi}=\delta_{ij}$.
By taking $n$ copies of \eqref{sc} at points $(\lambda,\mu)=(\lambda_{i},\mu
_{i})$ we obtain a linear system of $n$ separation
relations on $H_{k}$:
\begin{equation}
\lambda_i^{2n}+\sum_{k=1}^{n}H_{k}\lambda_i^{n-k}=\mu_i^{2},\qquad i=1,2,\ldots,n. \label{sr}%
\end{equation}
Solving \eqref{sr} with respect to $H_{k}$ yields $n$ functions
(Hamiltonians) in involution on the Poisson manifold $M$
\begin{equation}\label{sham}
H_{k}=\frac{1}{2}\bm{\mu}^{T}A_{k}G\bm{\mu}+V_{k}^{(2n)},\qquad k=1,\ldots,n.
\end{equation}
The matrix $G$ can be interpreted as a contravariant metric tensor on the
configuration space $Q$. It can be shown that the
metric $G$ is flat. Matrices $A_{k}$ (with $A_{1}=\mathbb{I}$) are $(1,1)$-Killing tensors for the
metric $G$ and $V_{k}^{(2n)}$ are the so-called basic
separable potentials. 

In what follows we will work in the so-called Viète (canonical) coordinates
$(\bm{q},\bm{p})\in T^{\ast}Q$ defined through
\begin{equation}
q_{i}=(-1)^{i}s_{i}(\bm{\lambda}),\qquad p_{i}=-\sum_{k=1}^{n}\frac{\lambda_{k}^{n-i}\mu_{k}}{\Delta_{k}},\qquad i=1,\ldots,n,\label{V1}
\end{equation}
in which the Stäckel Hamiltonians \eqref{sham} take the form
\begin{equation}
H_{k}=\frac{1}{2}\bm{p}^{T}A_{k}G\bm{p}+V_{k},\qquad k=1,\ldots,n, \label{Hk}
\end{equation}
where $\bm{p}=(p_{1},\ldots,p_{n})^{T}$ and $\bm{q}=(q_{1},\ldots,q_{n})^{T}$.
In \eqref{V1} $s_{i}(\bm{\lambda})$ are the elementary symmetric polynomials in $\lambda_{i}$ and
$\Delta_{j}=\prod\limits_{k\neq j}(\lambda_{j}-\lambda_{k})$. In Viète
coordinates all Hamiltonians \eqref{Hk} are polynomials of their arguments.
Explicitly
\[
G^{ij}=2q_{i+j-n-1},\qquad \left(A_{k}\right)_{j}^{i}=
\begin{cases}
q_{i-j+k-1}, & \text{$i\leqslant j$ and $k\leqslant j$}\\
-q_{i-j+k-1}, & \text{$i>j$ and $k>j$}\\
0, & \text{otherwise},%
\end{cases}
\]
where, for convenience, we set $q_{0}=1$ and $q_{k}=0$ for $k<0$ or $k>n$. 
In particular $(A_1)^i_j = \delta_i^j$.
The elementary separable potentials $V_{k}^{(\alpha)}$ can be constructed by the recursion formula
\cite{blaszak2011}
\[
  \bm{V}^{(\alpha)}=R^{\alpha}\bm{V}^{(0)},\qquad R=\begin{pmatrix}-q_{1} & 1 & 0 & 0\\
\vdots & 0 & \ddots & 0\\
\vdots & 0 & 0 & 1\\
-q_{n} & 0 & 0 & 0
\end{pmatrix}, 
\]
where $\bm{V}^{(\alpha)}=\bigl(V_{1}^{(\alpha)},\ldots,V_{n}^{(\alpha)}\bigr)^{T}$
and $\bm{V}^{(0)}=(0,\ldots,0,-1)^{T}$.

Thus, the separation curve \eqref{sc} generates the autonomous Hamiltonian system
\begin{equation}
\bm{\xi}_{t_{k}}= \{H_k,\bm{\xi}\}_\pi\equiv \bm{X}_{k},\qquad k=1,\ldots,n,\qquad n\in\mathbb{N}.\label{Hsys}%
\end{equation}
The first Hamiltonian
$H_{1}$ of this system can be interpreted as the Hamiltonian of a particle in
the pseudo-Riemannian $n$-dimensional configuration space $(Q,g=G^{-1})$
moving under particular potential force and the remaining Hamiltonians are its
constants of motion. By their very construction the Hamiltonians $H_{k}$ Poisson commute
\[
\{H_{i},H_{j}\}_{\pi}=\pi(dH_{i},dH_{j})=0,\qquad i,j=1,\ldots, n,%
\]
so that $[\bm{X}_{i},\bm{X}_{j}]=0$. It means that each system \eqref{Hsys} is Liouville
integrable and separable (in the sense of the Hamilton-Jacobi theory) Hamiltonian system. The coordinates $(\bm{\lambda},\bm{\mu})$ are by construction separation coordinates
of the system.

The Hamiltonian equations \eqref{Hsys} can be represented by isospectral Lax equations\cite{Blaszak2019a}
\begin{equation}
L_{t_k} = [U_{k}, L],\qquad
k=1,\ldots,n, \label{Hlax}%
\end{equation}
where the Lax matrix $L$ takes the form
\begin{equation}
L=\begin{pmatrix}v & u\\
w & -v
\end{pmatrix}\label{laxm}
\end{equation}
with the elements defined by
\begin{equation} \label{uv}
u=\lambda^{n}+\sum_{k=1}^{n}q_{k}\lambda^{n-k},\qquad v=-\frac{1}{2}\sum_{k=1}^{n}\Bigl[\sum_{i=1}^{n}G^{ki}p_{i}\Bigr]\lambda^{n-k} 
\end{equation}
and
\begin{equation} \label{uv2}
w = \left[\frac{\lambda^{2n}- v^2}{u}\right]_{+}.
\end{equation}
Here $\left[\frac{a(\lambda)}{b(\lambda)}\right]_{+}$ means the polynomial part
of the quotient of the polynomial $a(\lambda)$ and the polynomial $b(\lambda)$.
Due to the form of $w$ the spectral curve \eqref{sc}
can be reconstructed from the characteristic equation of the Lax matrix \eqref{laxm} since
\begin{equation}
0=\det\bigl[L-\mu\mathbb{I}\bigr]=-(v^{2}+uw)+\mu^{2}=-\Bigl(\lambda^{2n}+\sum_{k=1}^{n}H_{k}\lambda^{n-k}\Bigr)+\mu^{2},\label{T4}
\end{equation}
where the second equality in \eqref{T4} has been proved in \cite{Blaszak2019a}. Moreover, the auxiliary matrices $U_{k}$ are defined by
\begin{equation}
U_{k}:=\left[\frac{u_{k}L}{u}\right]_{+}\equiv\begin{pmatrix}\left[\frac{u_{k}v}{u}\right]_{+} & u_{k}\\
\left[\frac{u_{k}w}{u}\right]_{+} & -\left[\frac{u_{k}v}{u}\right]_{+}
\end{pmatrix},\qquad k=1,\ldots,n,\label{T8}
\end{equation}
where
\[
u_{k}:=\left[\frac{u}{\lambda^{n-k+1}}\right]_{+}\equiv\lambda^{k-1}+\sum_{i=1}^{k-1}q_{k}\lambda^{k-i-1}.
\]

One can show that, see \cite{2},\footnote{In fact, in \cite{2} we consider a slightly 
more general class of separation curves, for which in the associated Lax representations there appears an important polynomial function $Q$, which
for the separation curves of the form \eqref{sc} reduces to
\[Q=-\left[\frac{w}{u}\right]_{+}=
-\left[\frac{\lambda^{2n}}{u^2}\right]_{+} = -1\]
by \eqref{uv} and \eqref{uv2}, cf.~(85) and (86) in \cite{2}.
}
\[
\dot{u}=-2v,\qquad\dot{v}=w-u.
\]
Here again the dot means the derivative with respect
to $t_1$. Hence, we can rewrite the Lax matrix \eqref{laxm} as
\begin{equation}
L=\begin{pmatrix}-\frac{1}{2}\dot{u} & u\\
-\frac{1}{2}\ddot{u}+u & \frac{1}{2}\dot{u}
\end{pmatrix},\label{T11}
\end{equation}
and the auxiliary matrices \eqref{T8} as
\begin{equation}
U_{k}=\begin{pmatrix}
-\frac{1}{2}\dot{u}_{k} & u_{k}\\
-\frac{1}{2}\ddot{u}_{k}+u_{k} & \frac{1}{2}\dot{u}_{k}
\end{pmatrix},\qquad k=1,\ldots,n.\label{T12}
\end{equation}
Then, the characteristic equation \eqref{T4} can be given in the form
\[
-\frac{1}{2}u\ddot{u}+\frac{1}{4}\dot{u}^{2}+u^{2}=\mu^{2}.
\]
Moreover, using \eqref{T11} and \eqref{T12} we see that the Lax equation \eqref{Hlax} for
$k=1$ yields the following ODE
\begin{equation}
\dddot{u}-4\dot{u}=0,\label{T14}
\end{equation}
while the remaining (i.e.~for $k=2,\ldots,n$) Lax equations \eqref{Hlax}
yield, due to \eqref{T14}, the following hierarchy of PDE's
\[
u_{t_{k}} =\dot{u}u_{k}-u\dot{u}_{k},\qquad
\dddot{u}_{k}-4\dot{u}_{k} = 0.
\]

\begin{example}\label{eH1} 
Consider the separation curve \eqref{sc} for $n=2$. 
The two associated Stäckel Hamiltonians \eqref{Hk} are given in Viète coordinates by 
\begin{align}\label{Hn2}
H_{1} =2p_{{1}}p_{{2}}+q_{{1}}\,{p_{{2}}^{2}+q}_{1}^{3}-2q_{1}%
q_{2},\quad
H_{2}  ={p_{{1}}^{2}}+2q_{{1}}p_{{1}}p_{{2}}+\left(  {q_{{1}}^{2}}-q_{{2}%
}\right)  {p_{{2}}^{2}+q}_{1}^{2}q_{2}-q_{2}^{2}.
\end{align}%
They generate Stäckel system \eqref{Hsys} that consists of
the two following evolution equations
\[
\begin{pmatrix}
q_{1}\\ q_{2}\\ p_{1}\\ p_{2}%
\end{pmatrix}_{t_1}
=
\begin{pmatrix}
2p_{2}\\ 2q_{1}p_{2}+2p_{1}\\ -p_{2}^{2}-3q_{1}^{2}+2q_{2}\\ 2q_{1}%
\end{pmatrix}\equiv \bm{X}_1,\quad
\begin{pmatrix}
q_{1}\\ q_{2}\\ p_{1}\\ p_{2}%
\end{pmatrix}_{t_2}
=
\begin{pmatrix}
2p_{1}+2q_{1}p_{2}\\ 2q_{1}p_{1}+2\left(  {q_{{1}}^{2}}-q_{{2}}\right)  {p_{{2}}}\\
-2p_{1}p_{2}-2q_{1}p_{2}^{2}-2q_{1}q_{2}\\ p_{2}^{2}-q_{1}^{2}+2q_{2}%
\end{pmatrix}\equiv \bm{X}_2.
\]
The Lax representation \eqref{Hlax} is provided by the Lax matrix
\[
L= \begin{pmatrix}
  -p_{2}\lambda-q_{1}p_{2}-p_{1} & \lambda^{2}+q_{1}\lambda+q_{2}\\
\lambda^{2}-q_{1}\lambda-p_{2}^{2}+q_{1}^{2}-q_{2} & p_{2}\lambda+q_{1}p_{2}+p_{1}%
\end{pmatrix}
\]
and the auxiliary matrices
\[
U_{1}=
\begin{pmatrix}
  0 & 1\\ 1 & 0
\end{pmatrix},\qquad 
U_{2}=
\begin{pmatrix}
  -p_{2} & \lambda+q_{1}\\ \lambda-q_{1} & p_{2}
\end{pmatrix}.
\]
\end{example}

\begin{example}\label{eH2} 
For the separation curve \eqref{sc} for $n=3$ 
the Stäckel Hamiltonians \eqref{Hk} in Viète coordinates are
\begin{equation}\label{Hn3}
\begin{split}
H_{1} &=   2p_{1}p_{3}+p_{2}^{2}+2q_{1}p_{2}p_{3}+q_{2}p_{3}^{2}
-q_{1}^{4}+3q_{1}^{2}q_{2}-2q_{1}q_{3}-q_{2}^{2},\\
H_{2} &=  2q_{1}p_{1}p_{3}+2q_{1}p_{2}^{2}+2p_{1}p_{2}+(q_{1}q_{2}-q_{3})p_{3}^{2}
+2q_{1}^{2}p_{2}p_{3} -q_{1}^{3}q_{2}+q_{1}^{2}q_{3}\\
&\quad +2q_{1}q_{2}^{2}-2q_{2}q_{3},\\
H_{3} &= p_{1}^{2}+2q_{1}p_{1}p_{2}+2q_{2}p_{1}p_{3}+q_{1}^{2}p_{2}^{2}+(q_{2}^{2}
-q_{1}q_{3})p_{3}^{2}+2(q_{1}q_{2}-q_{3})p_{2}p_{3} -q_{1}^{3}q_{3}\\
&\quad +2q_{1}q_{2}q_{3}-q_{3}^{2}.
\end{split}
\end{equation}
The related evolution equations \eqref{Hsys} are%
\[
\begin{pmatrix}
q_{1}\\ q_{2}\\ q_3\\ p_{1}\\ p_{2}\\ p_3%
\end{pmatrix}_{t_1}
=
\begin{pmatrix}
2p_{3}\\ 2q_{1}p_{3}+2p_{2}\\ 2q_{1}p_{2}+2q_{2}p_{3}+2p_{1}\\
-2p_{2}p_{3}+4q_{1}^{3}-6q_{1}q_{2}+2q_{3}\\ -p_{3}^{2}-3q_{1}^{2}+2q_{2}\\ 2q_{1}%
\end{pmatrix}\equiv \bm{X}_1,
\]
\[
\begin{pmatrix}
q_{1}\\ q_{2}\\ q_3\\ p_{1}\\ p_{2}\\ p_3%
\end{pmatrix}_{t_2}
=
\begin{pmatrix}
2p_{2}+2q_{1}p_{3}\\ 2p_{1}+2q_{1}^{2}p_{3}+4q_{1}p_{2}\\
2q_{1}p_{1}+2(q_{1}q_{2}-q_{3})p_{3}+2q_{1}^{2}p_{2}\\
-2p_{1}p_{3}-2p_{2}^{2}-4q_{1}p_{2}p_{3}-q_{2}p_{3}^{2}+3q_{1}^{2}q_{2}-2q_{1}q_{3}-2q_{2}^{2}\\
-q_{1}p_{3}^{2}+q_{1}^{3}-4q_{1}q_{2}+2q_{3}\\
p_{3}^{2}-q_{1}^{2}+2q_{2}
\end{pmatrix}\equiv \bm{X}_2,
\]
\[
\begin{pmatrix}
q_{1}\\ q_{2}\\ q_3\\ p_{1}\\ p_{2}\\ p_3%
\end{pmatrix}_{t_3}
=
\begin{pmatrix}
2q_{1}p_{2}+2q_{2}p_{3}+2p_{1}\\ 2q_{1}^{2}p_{2}+2q_{1}p_{1}+2(q_{1}q_{2}-q_{3})p_{3}\\
2q_{2}p_{1}+2(q_{2}^{2}-q_{1}q_{3})p_{3}+2(q_{1}q_{2}-q_{3})p_{2}\\
-2p_{1}p_{2}-2q_{1}p_{2}^{2}-2q_{2}p_{2}p_{3}+q_{3}p_{3}^{2}-2q_{2}q_{3}+3q_{1}^{2}q_{3}\\
-2p_{1}p_{3}-2q_{1}p_{2}p_{3}-2q_{2}p_{3}^{2}-2q_{1}q_{3}\\
q_{1}p_{3}^{2}+2p_{2}p_{3}+q_{1}^{3}+2q_{3}-2q_{1}q_{2}%
\end{pmatrix}\equiv \bm{X}_3.
\]
The associated Lax equations \eqref{Hlax} are given by
\[
L=
\begin{pmatrix}
  -p_{3}\lambda^{2}-(q_{1}p_{3}+p_{2})\lambda-p_{1}-q_{1}p_{2}-q_{2}p_{3} &
  \lambda^{3}+q_{1}\lambda^{2}+q_{2}\lambda+q_{3}\\
  \lambda^{3}-q_{1}\lambda^{2}+(q_{1}^{2}-q_{2}-p_{3}^{2})\lambda + \star
  &
  p_{3}\lambda^{2}+(q_{1}p_{3}+p_{2})\lambda+p_{1}+q_{1}p_{2}+q_{2}p_{3}
\end{pmatrix},
\]
where
$\star =  -q_{1}p_{3}^{2}-2p_{2}p_{3}-q_{1}^{3}+2q_{1}q_{2}-q_{3}$, and
\[
U_{1} =
\begin{pmatrix}
  0 & 1\\ 1 & 0
\end{pmatrix},\qquad
U_{2}=
\begin{pmatrix}
  -p_{3} & \lambda+q_{1}\\ \lambda-q_{1} & p_{3}
\end{pmatrix},
\]
\[
U_{3}=
\begin{pmatrix}
  -p_{3}\lambda-p_{2}-q_{1}p_{3} & \lambda^{2}+q_{1}\lambda+q_{2}\\
\lambda^{2}-q_{1}\lambda+q_{1}^{2}-q_{2}-p_{3}^{2} & p_{3}\lambda+p_{2}%
+q_{1}p_{3}
\end{pmatrix}.
\]
\end{example}

\section{Stäckel representation of the AKNS stationary systems\label{s4}}

In order to relate the stationary AKNS system \eqref{statsys} with
a corresponding Stäckel system \eqref{Hsys}, associated with the spectral curve \eqref{sc}, we will need to make an identification between their Lax representations \eqref{Laxs} and \eqref{Hlax}. Let us note that the
degrees of $\lambda$-polynomials in the elements of the auxiliary matrices \eqref{aux} constrained by $\bm{K}_{n+1}=0$ and the corresponding auxiliary matrices \eqref{T8}
do not agree. However, for a given Lax representation of the AKNS hierarchy
\[
\mathbb{L}_{t_k}=[\mathbb{U}_k,\mathbb{L}]+(\mathbb{U}_k)_x, \quad k=1,\ldots,n,
\]
there exist infinitely many gauge equivalent Lax representations
\[
\mathbb{L}'_{t_k}=[\mathbb{U}'_k,\mathbb{L}'] + (\mathbb{U}'_k)_x, \quad k=1,\ldots,n,
\]
where
\[
\mathbb{L}^{\prime}=\Omega \mathbb{L}\Omega^{-1},\qquad \mathbb{U}'_{k}=\Omega \mathbb{U}_k \Omega^{-1},  
\]
with $\Omega$ being any $2\times 2$ constant invertible matrix. Each such gauge transformation induces 
the following gauge transformation of the Lax representation \eqref{Laxs} of the corresponding stationary AKNS system \eqref{statsys}:
\[
(\mathbb{U}_{n+1})_{t_k}=[\mathbb{U}_k,\mathbb{U}_{n+1}]\qquad \Longleftrightarrow\qquad (\mathbb{U}'_{n+1})_{t_k}=[\mathbb{U}'_k,\mathbb{U}'_{n+1}].
\]
Obviously, the associated spectral curve \eqref{L11} is preserved under this gauge transformation as
\[
\det \left( \mathbb{U}_{n+1}-\mu \mathbb{I}\right ) = \det \left( \mathbb{U}'_{n+1}-\mu \mathbb{I}\right ).  
\]

In our case, the unique gauge transformation $\mathbb{U}_k\mapsto \mathbb{U}'_k$
equalizing polynomial degrees of matrix elements of $\mathbb{U}'_{k}$ \eqref{aux} and $\mathbb{U}_{k}$ \eqref{T8},  
with an appropriate sign of the leading terms, is the rotation by $-\frac{1}{4}\pi$. Thus,
for 
\[
\Omega = \begin{pmatrix}
  \frac{1}{\sqrt{2}} & \frac{1}{\sqrt{2}}\\
-\frac{1}{\sqrt{2}} & \frac{1}{\sqrt{2}}
\end{pmatrix} 
\]
we have 
\begin{equation}\label{gL}
  \mathbb{L}'= \Omega \mathbb{L} \Omega^{-1} = \begin{pmatrix}
\frac{1}{2}(q+r) & \lambda+\frac{1}{2}(q-r)\\
\lambda-\frac{1}{2}(q-r) & -\frac{1}{2}(q+r)
\end{pmatrix} \equiv \mathbb{U}'_2
\end{equation}
and for $k\geqslant 1$
\begin{equation}\label{gUk}
  \mathbb{U}'_k = \Omega \mathbb{U}_k \Omega^{-1} =
\frac{1}{2}\begin{pmatrix}
  \mathbb{B}_k+\mathbb{C}_k & -2\mathbb{A}_k + \mathbb{B}_k-\mathbb{C}_k\\
  -2\mathbb{A}_k - \mathbb{B}_k-\mathbb{C}_k  & \mathbb{B}_k+\mathbb{C}_k
\end{pmatrix}  \equiv 
\begin{pmatrix}
-\frac{1}{4}\dot{\mathbb{P}}_{k} & \frac{1}{2}\mathbb{P}_{k}\\
-\frac{1}{4}\ddot{\mathbb{P}}_{k}+\frac{1}{2}\mathbb{P}_{k} & \frac{1}{4}\dot{\mathbb{P}}_{k}
\end{pmatrix}.
\end{equation}

By comparing \eqref{gL} with \eqref{T11} and \eqref{gUk} with \eqref{T12} we can now see that the Lax representations of the $n$-th stationary AKNS system \eqref{statsys}
and the Stäckel system \eqref{Hsys} associated with the spectral curve \eqref{sc} coincide provided that
\begin{equation}\label{idn}
  L \equiv \mathbb{U}'_{n+1}\quad \text{or equivalently}\quad  u\equiv \frac{1}{2}\mathbb{P}_{n+1}
  =-\mathbb{A}_{n+1} + \frac{1}{2}\mathbb{B}_{n+1}-\frac{1}{2}\mathbb{C}_{n+1}.
\end{equation} 
Notice that, in both cases, the auxiliary matrices
are completely determined by \eqref{idn}, since we have the equalities \eqref{Un} 
and \eqref{T8}.\footnote{Notice that from \eqref{Un} it follows that 
$\mathbb{U}_k = \left[\frac{\mathbb{U}_{n+1}}{\lambda^{n-k+1}}\right]_{+}$, where $k=1,\ldots,n$,
and compare it with \eqref{T8}.} Thus, we have also the identifications
\[
  U_k\equiv \mathbb{U}'_k\quad\text{and}\quad u_k\equiv \frac{1}{2}\mathbb{P}_k
  =-\mathbb{A}_{k} + \frac{1}{2}\mathbb{B}_{k}-\frac{1}{2}\mathbb{C}_{k},\qquad k=1,\ldots, n,
\]
defined by the auxiliary matrices \eqref{T12} and \eqref{gUk}, respectively. Furthermore, 
\[
    v = -\frac{1}{2}\dot{u}\equiv -\frac{1}{4}\dot{\mathbb{P}}_{n+1} 
    = \frac{1}{2} \mathbb{B}_{n+1} + \frac{1}{2}\mathbb{C}_{n+1}.
\]
Hence, taking into account \eqref{uv} we arrive at the main result of this paper.
\begin{theorem}\label{th}
The transformation between jet coordinates 
$q,r,q_x,r_x,\ldots,\linebreak[0]q_{(n-1)x},r_{(n-1)x}$ on the stationary
manifold $\mathcal{M}_{n}\equiv M_n$ and Viète coordinates
$(\bm{q},\bm{p})$, given by 
\begin{equation}\label{map1}
q_{i}=\frac{1}{2}P_{i} =  -\frac{1}{2}a_i - \frac{1}{4}b_i - \frac{1}{4}c_i,\qquad
 p_{i}=\frac{1}{2}\sum_{j=1}^{n}\bigl(G^{-1}\bigr)_{ij}(b_j-c_j),\qquad i=1,\ldots,n, 
\end{equation}
transforms the $n$th stationary AKNS system \eqref{statsys} to the Stäckel system  \eqref{Hsys} given by the separation curve \eqref{sc}.
\end{theorem}
Using \eqref{K12} we can write the map \eqref{map1} as
\begin{equation}\label{map2}
\begin{split}
q_{i}&=-\frac{1}{4}(K_i)_1-\frac{1}{4}(K_i)_2 - \frac{1}{2}\partial_x^{-1}\left [ r (K_i)_1 + q (K_i)_2\right],\\ 
p_{i}&=\frac{1}{2}\sum_{j=1}^{n}\bigl(G^{-1}\bigr)_{ij}\left[ (K_j)_1-(K_j)_2\right],
\end{split}
\end{equation}
where $i=1,\ldots,n$. In the above maps the metric $G^{-1}$ is expressed in the coordinates on the configuration 
space provided by the first part of the map \eqref{map1} or \eqref{map2}. Also, under this map, we have $H_k\equiv -h_{n-k}$, $k=1,\ldots,n$.

\begin{example}
Let $n=2$. The transformation \eqref{map2} on the stationary manifold $\mathcal{M}_2$ from jet to Viète coordinates is
\begin{align*}
q_{1}&=\frac{1}{2}(q-r),\qquad q_{2}=-\frac{1}{4}q_{x}-\frac{1}{4}r_{x}-\frac{1}{2}qr,\\
p_{1}&=\frac{1}{4}(q_{x}-r_{x})+\frac{1}{4}(q^{2}-r^{2}),\qquad p_{2}=-\frac{1}{2}(q+r),
\end{align*}
and its inverse is
\begin{align*}
 q &= q_1-p_2,\qquad r= -(q_1+p_2),\\
 q_x &= 2 p_1+2 q_1 p_2-p_2^2+q_1^2-2 q_2,\qquad r_x = -2 p_1-2 q_1 p_2-p_2^2+q_1^2-2 q_2. 
\end{align*}
It maps Stäckel Hamiltonians \eqref{Hn2} onto constants of motion \eqref{hn2}, so that
$H_1 = - h_1$ and $H_2 = -h_0$, and the associated vector fields $\bm{K}_1, \bm{K}_2$
from Example \ref{ex1} to Hamiltonian vector fields $\bm{X}_1, \bm{X}_2$ from Example \ref{eH1}, respectively.
\end{example}

\begin{example}
Let $n=3$, then the transformation \eqref{map2} on the stationary manifold $\mathcal{M}_3$ is
\begin{align*}
q_{1}  &  =\frac{1}{2}(q-r),\qquad 
q_{2}=-\frac{1}{4}q_{x}-\frac{1}{4}r_{x}-\frac{1}{2}qr,\\
q_{3} &=\frac{1}{4}rq_{x}-\frac{1}{4}qr_{x}+\frac{1}{8}q_{xx}
-\frac{1}{8}r_{xx}+\frac{1}{4}qr^{2}-\frac{1}{4}q^{2}r,\\
p_{1}  &  =-\frac{1}{8}(q_{xx}+r_{xx})-\frac{1}{4}(rr_{x}+qq_{x})
+\frac{1}{8}(qr^{2}+q^{2}r-r^{3}- q^{3})\\
p_{2}  &  =\frac{1}{4}(r_{x}-q_{x})+\frac{1}{4}(q^{2}-r^{2}),\qquad 
p_{3}=-\frac{1}{2}(q+r),
\end{align*}
and its inverse is
\begin{align*}
 q&= q_1-p_3,\qquad q_x= 2 p_3 q_1-p_3^2+2 p_2+q_1^2-2 q_2,\\
 q_{xx}&= 4 p_3^2 q_1+2 p_3 q_1^2-4 p_3 q_2-4 p_2 q_1-2 p_3^3+4 p_2 p_3-4 p_1-4 q_1 q_2+4 q_3,\\
 r&= -p_3-q_1,\qquad r_x = -2 p_3 q_1-p_3^2-2 p_2+q_1^2-2 q_2,\\
r_{xx}&= -4 p_3^2 q_1+2 p_3 q_1^2-4 p_3 q_2-4 p_2 q_1-2 p_3^3-4 p_2 p_3-4 p_1+4 q_1 q_2-4 q_3. 
\end{align*}
This transformation maps Stäckel Hamiltonians \eqref{Hn3} onto constants of motion \eqref{hn3}, so that
$H_1 = - h_2, H_2 = - h_1$ and $H_3 = -h_0$, and the vector fields $\bm{K}_1, \bm{K}_2$
and $\bm{K}_3$ from Example \ref{ex2} to Hamiltonian vector fields $\bm{X}_1, \bm{X}_2$ and $\bm{X}_3$ 
from Example \ref{eH2}, respectively.
\end{example}

\end{document}